\documentclass[journal=jctcce,manuscript=article]{achemso}

\usepackage[utf8x]{inputenc}
\usepackage[T1]{fontenc}
\usepackage{lmodern}
\usepackage{amsmath, amssymb}
\usepackage{mathrsfs}
\usepackage{braket}
\usepackage{booktabs}
\usepackage{graphicx}
\usepackage{array, dcolumn, makecell, multirow}
\usepackage{siunitx}
\usepackage[version = 3]{mhchem}
\usepackage[printonlyused, withpage]{acronym}
\usepackage{paralist}
\usepackage[colorlinks = true, linkcolor = blue, citecolor = blue, urlcolor = blue, bookmarks = true]{hyperref}
\usepackage[normalem]{ulem}

\title{Dispersion energy of symmetry-adapted perturbation theory from explicitly correlated F12 approach}
\author{Michał Przybytek}
\email{michal.przybytek@tiger.chem.uw.edu.pl}
\affiliation{Faculty of Chemistry, University of Warsaw, ul.~L.~Pasteura 1, 02-093 Warsaw, Poland}

\begin{document}

\begin{abstract}
Methods of the explicitly correlated F12 approach are applied to the problem of calculating the uncoupled second-order dispersion energy in symmetry-adapted perturbation theory. 
The accuracy of the new method is tested for noncovalently bound complexes from the A24 data set [J. \v{R}ez\'{a}\v{c} and P. Hobza, J. Chem. Theory Comput. {\bf 9}, 2151 (2013)] using standard orbital basis sets aug-cc-pV$X$Z supplemented with auxiliary aug-cc-pV$X$Z\_OPTRI sets. 
For near equilibrium geometries, it is possible to recover the dispersion energy with average relative errors consistently smaller than 0.1\% (with respect to the CBS extrapolated limit estimated from regular orbital calculations). 
This level of accuracy is achieved already in basis set of a triple-zeta quality, when a Slater-type correlation factor $\exp(-0.9\,r_{12})$ is combined with variant C of the F12 approach.
The explicitly correlated approach clearly outperforms regular orbital calculations in the basis set of 5-zeta quality (average relative errors of 1\%).
\end{abstract}

\maketitle

\section{Introduction}

Symmetry-adapted perturbation theory (SAPT) is one of the most successful \textit{ab initio} methods capable of describing noncovalent interactions.~\cite{Jeziorski:94,Jeziorski:03,Szalewicz:12,Hohenstein:12,Jansen:14} Among the main strengths of SAPT is the possibility to interpret the nature of intermolecular forces---the expansion of the interaction energy in terms of the intermolecular interaction operator provides decomposition into well-defined energy contributions. In low orders they are electrostatic, exchange, induction and dispersion energies. Continuous effort to extend the applicability and accuracy of SAPT resulted in development of wave function-based\cite{Szalewicz:79,Jeziorski:94,Hohenstein:12} and Kohn-Sham-based\cite{Misquitta:05,Hesselmann:05,Jansen:14} formulations capable of treating large molecular systems.~\cite{Hohenstein:12,Hesselmann:14} When supplemented with approximate treatment of higher-order energy corrections, SAPT may rival the accuracy of best-performing supermolecular methods.~\cite{Hohenstein:12,Parker:14,Lao:15}

The dispersion energy accounts for the intermolecular correlation effects and its accurate description in weakly interacting systems still poses a significant challenge for \textit{ab initio} methods.~\cite{Grimme:16,Hermann:17} It is well-known that the basis set convergence of dispersion energy is particularly slow which follows from the necessity to model the interelectronic cusp of the wave function in regions between the molecules. Standard finite expansion of the wave function in terms of orbital products is a poor choice as the number of atom-centered basis functions required to saturate dispersion quickly becomes prohibitively large.

There exists several strategies to overcome the slow basis set convergence of electron correlation energy with respect to the basis set size in standard orbital-product expansions: extrapolation to the complete basis set (CBS) limit,~\cite{Halkier:98,Schwenke:05} extending the basis set with functions centered in regions between the molecules (midbond functions),\cite{Tao:91,Gutowski:87,Williams:95} or a combination of the two.~\cite{Jeziorska:03,Patkowski:12} It is also possible to reduce the computational effort by employing Gaussian basis sets dedicated to intermolecular interactions in which exponents were optimized with respect to dispersion energy.~\cite{Williams:95,Bukowski:99,Przybytek:05,Patkowski:10}

In SAPT both midbond functions and CBS extrapolation techniques were applied to saturate the dispersion energy contribution. The former may be added either in monomer-centered or dimer-centered SAPT calculations yielding the so-called MC$^+$BS and DC$^+$BS approaches, respectively.~\cite{Williams:95} Regarding the CBS extrapolation of dispersion energy, a popular choice in the SAPT community are two-point extrapolation schemes developed for electron correlation energies.~\cite{Hesselmann:06,Sedlak:10,Sirous:15,Sirous:16} One should mention the work of \v{R}ez\'{a}\v{c} and Hobza\cite{Rezac:11} who used the Schwenke's power-law formula\cite{Schwenke:05} to derive a two-point extrapolation scheme for the dispersion energy calculated at the DFT-SAPT level of theory. Additionally, they proposed scaling factors for one-point schemes. Scaling factors may also be used on top of the two-point extrapolation, e.g. see recent DFT-SAPT study of Hesselmann and Korona.~\cite{Hesselmann:14}  

A viable alternative for improving the basis set convergence is to use wave functions that include terms with explicit dependence on the interelectronic distance.~\cite{Helgaker:08,Hattig:11,Kong:12} This leads to much better description of the correlation (Coulomb) hole---the decline of the wave function amplitude as the electrons approach each other.~\cite{Prendergast:01,Szalewicz:10} In fact, for few-electron atoms and small molecules explicitly correlated methods reach unrivalled accuracy that allows \textit{ab initio} calculations to contribute in establishing new metrology standards and future International System of Units (SI).~\cite{Cencek:12,Piszczatowski:15,Moldover:16,Fischer:16,Przybytek:17}


The concept of explicit correlation was introduced in SAPT already in the early days of the method when explicitly correlated Gaussian-type geminals (GTG) were employed in calculations for the helium dimer.~\cite{Chalbie:77,Szalewicz:79} For this system it is possible to express with GTG expansion the first-order\cite{Rybak:89} as well as the dominant second-order energy contributions.~\cite{Williams:96,Korona:97,Jeziorska:07} The latter include dispersion and exchange-dispersion terms: $E^{(20)}_\mathrm{disp}$, $E^{(21)}_\mathrm{disp}$ and $E^{(20)}_\mathrm{exch-disp}$, where the upper $(ij)$ index refers to the $i$th and $j$th order corrections with respect to the intermolecular interaction operator and intramolecular correlation operator, respectively. The combination of geminal basis for accurate treatment of low-level corrections and large Gaussian-type orbital basis for higher order SAPT components led to highly-accurate nonrelativistic Born-Oppenheimer potentials for the helium dimer.~\cite{Korona:97,Jeziorska:07} The computational cost of the GTG-based SAPT approach is high due to the need for optimization of a large number of nonlinear parameters in primitive basis functions, and evaluation of three- and four-electron integrals.~\cite{Szalewicz:10} In consequence helium dimer remains the only system treated with this method.

In 1985 the work of Kutzlenigg\cite{Kutzelnigg:85} marked the advent of a new class of explicitly correlated approaches known as R12 methods\cite{Werner:07,Helgaker:08,Hattig:11,Ten-no:12,Kong:12} developed later by Kutzelnigg, Klopper and Noga.~\cite{Klopper:87,Klopper:91,Kutzelnigg:91,Noga:92} The key idea is to construct pair functions from the expansion in products of occupied Hartree-Fock orbitals multiplied by a single linear $r_{12}$ term, in addition to regular products of all virtual orbitals. Therefore, pair functions depend on a universal correlation factor which is not optimized, yet fulfills the cusp condition. Another crucial step is to factorize the resulting three- and four-electron integrals into sums of products of two-electron integrals by using the resolution of identity (RI). The RI approximation is most efficient when carried out in separate auxiliary basis sets, as first shown by Klopper and Samson,~\cite{Klopper:02} or with complementary auxiliary basis set (CABS approach).~\cite{Valeev:04}

Although the linear correlation factor in the original R12 approach introduces the cusp in the wave function and is valid for short electron-electron distances, it simultaneously fails to restore the correct behavior of the wave function when the electrons are far apart. The first solution to ameliorate this was the orbital invariant formulation proposed by Klopper.~\cite{Klopper:91} The second possibility is to replace the linear correlation factor with a more flexible $f(r_{12})$ function which more accurately describes the Coulomb hole. Nowadays, the most popular choice is the short-range Slater-type correlation factor $[1-\exp(-\gamma r_{12})]/\gamma$ introduced by Ten-no,~\cite{Ten-no:04} yet efforts to improve the form of the correlation factor continue.~\cite{Tew:05,Lesiuk:13,Silkowski:15} The generalization to arbitrary $f(r_{12})$ correlation factors is referred to as the F12 formalism.~\cite{Manby:04}

Due to approximations which avoid many-electron integrals and nonlinear optimization the R12/F12 variants of MP2 and CC methods are affordable even for large polyatomic systems with only marginal computational overhead with respect to the parent approach. At the same time, the convergence of the correlation energy with the basis set size is significantly improved. For instance, CCSD(T)-F12 results obtained in small augmented correlation-consistent double-zeta basis of Dunning\cite{Dunning:89,Kendall:92} are comparable to the conventional CCSD(T) results of quadruple- or quintuple-zeta quality.~\cite{Hattig:11}

Until now, the combination of SAPT and explicitly correlated F12 methods has not been a direct one. Specifically, several corrections based on results of explicitly correlated calculations have been proposed to compensate for the basis set incompleteness of the SAPT dispersion energy. In the study of imidazole-benzene and pyrrole-benzene complexes\cite{Ahnen:14} second-order dispersion and exchange-dispersion energy components taken from DFT-SAPT were simply scaled by the ratio of CCSD(F12)+(T*) and CCSD(T) electron-correlation energies. Frey \textit{et al.} proposed an alternative approach, termed SAPT-F12(MP2),~\cite{Frey:16,Holzer:17} which amends DFT-SAPT results with the correlation contribution from MP2-F12. The introduced $\Delta$F12 correction incorporates all second-order pair energies in which two occupied orbitals are localized on different molecules. A more rigorous approach based on explicitly correlated ring-coupled-cluster-doubles method has also recently been suggested.~\cite{Hehn:16}

The aim of this work is to present the first direct and rigorous application of the F12 framework to calculations of the second-order dispersion energy  at the uncoupled level of theory, i.e. neglecting intramolecular correlation effects. The new method will be denoted $E^{(20)}_\mathrm{disp}$(F12). We wish to verify whether approximations applied in the modern F12 methods allow for perturbative description of the intermolecular correlation effects at a similar level of accuracy as may be achieved for electron correlation effects within a single molecule. To this end, we apply only the essential set of approximations necessary to avoid calculation of more than two-electron integrals. In particular, we do not use the extended Brillouin condition (EBC) in derivation of the formulas, and do not employ density fitting techniques in calculations. We examine different forms of the correlation factor and verify whether the existing auxiliary basis sets for the RI approximation can be used in a straightforward manner in the new context.

This manuscript is organized as follows. In Section~\ref{sec:theory} we derive equations of the $E^{(20)}_\mathrm{disp}$(F12) method, and present details of its computer implementation in Section~\ref{sec:details}. In Section~\ref{sec:results} we discuss numerical results for systems of the A24 data set.~\cite{Rezac:13} Section~\ref{sec:conclusions} concludes the paper and offers perspectives for future developments.


\section{\label{sec:theory}Theory}

\subsection{Orbital spaces}

In the following we will consider interaction between two closed-shell molecules $A$ and $B$ in their ground states. By solving the Hartree-Fock equations in a finite orbital basis set, separately for both molecules $A$ and $B$, we obtain two sets of orthonormal molecular orbitals (MO) and corresponding orbital energies
\begin{equation}
F_A\varphi^A_p=\varepsilon^A_p\varphi^A_p, \quad 
F_B\varphi^B_q=\varepsilon^B_q\varphi^B_q.
\end{equation}

After splitting the set of MOs of a given molecule into sets of occupied and virtual orbitals we introduce one-electron projection operators onto the occupied, virtual, and full MO spaces. For molecule $A$ the projectors are defined as
\begin{align}
\label{eq:projO}
O_A&=\sum_{i\in \mathrm{occ}(A)}|\varphi^A_i\rangle\langle\varphi^A_i|,\\
\label{eq:projV}
V_A&=\sum_{a\in \mathrm{virt}(A)}|\varphi^A_a\rangle\langle\varphi^A_a|,\\
\label{eq:projP}
P_A&=(O+V)_A=\sum_{p\in \mathrm{MO}(A)}|\varphi^A_p\rangle\langle\varphi^A_p|.
\end{align}
Following the CABS approach,~\cite{Valeev:04} the approximate RI expansion is performed in a union of the initial basis set (the one used to construct the MOs) and a complementary auxiliary basis set. With this choice, it is always possible to construct an orthonormal basis of auxiliary molecular orbitals (AMO) which can be decomposed into the initial MO space and its orthogonal complement, denoted by cmpl.~\cite{Valeev:04,Werner:07} This AMO set is used to approximate the projector $(1-P)_A$ and the identity operator
\begin{align}
\label{eq:projX}
(1-P)_A&\approx \tilde{X}_A=\sum_{a'\in \mathrm{cmpl}(A)}|\varphi^A_{a'}\rangle\langle\varphi^A_{a'}|,\\
\label{eq:proj1}
1_A&\approx\tilde{1}_A=\sum_{p'\in \mathrm{AMO}(A)}|\varphi^A_{p'}\rangle\langle\varphi^A_{p'}|.
\end{align}

The projectors $O_B$, $V_B$, $P_B$, $\tilde{X}_B$, and $\tilde{1}_B$ are defined in a similar way but in terms of orbitals obtained from calculations for molecule $B$. The system of naming orbital indices from various orbital spaces for both molecules is summarized in Table~\ref{tab:index}.

\begin{table}
\caption{\label{tab:index} Naming convention for orbital indices}
\begin{tabular}{lcc}
orbital space & molecule $A$ & molecule $B$ \\
\hline
occupied & $i,k,m,x,z$ & $j,l,n,y,w$ \\
virtual  & $a,c$   & $b,d$   \\
MO       & $p,r$   & $q,s$   \\
complementary & $a',c'$ & $b',d'$ \\
AMO & $p',r'$ & $q',s'$ \\
\end{tabular}
\end{table}

\subsection{Variational formulation of the dispersion energy}

To simplify notation, from now on we assume that all functions and operators labeled by $A$ and $B$ depend on coordinates of electron 1 and 2, respectively. Furthermore, a sum of the Fock operators for molecule $A$ and $B$ will be denoted by $F_{AB}$, that is
\begin{equation}
\label{eq:defAB}
F_{AB}=F_A+F_B.
\end{equation}

In the SAPT approach, the dispersion energy $E^{(20)}_\text{disp}$ can be expressed as a sum of pair contributions
\begin{equation}
E^{(20)}_\text{disp}=\sum_{ij}e_{ij}.
\end{equation}
The individual pair energies obey the variational condition~\cite{Jeziorski:76,Chalasinski:76}
\begin{equation}
\label{eq:paire}
e_{ij}\leq \varepsilon_{ij}=4\langle\varphi^A_i\varphi^B_j |r_{12}^{-1}\tau^{(ij)}\rangle,
\end{equation}
where the pair functions $\tau^{(ij)}$ are obtained by minimizing the functional
\begin{equation}
\label{eq:functional}
J_{ij}[\tilde{\tau}^{(ij)}]=\langle\tilde{\tau}^{(ij)}|F_{AB}-\varepsilon^A_i-\varepsilon^B_j|\tilde{\tau}^{(ij)}\rangle
+2\langle\tilde{\tau}^{(ij)}|r_{12}^{-1}\varphi^A_i\varphi^B_j\rangle,
\end{equation}
with trial functions $\tilde{\tau}^{(ij)}(1,2)$ satisfying the strong orthogonality (SO) conditions
\begin{align}
\label{eq:soA}
\langle\tilde{\tau}^{(ij)}|\varphi^A_k\xi\rangle=0&\quad \text{all}\;k,\;\text{any}\;\xi(2),\\
\label{eq:soB}
\langle\tilde{\tau}^{(ij)}|\zeta\varphi^B_l\rangle=0&\quad \text{all}\;l,\;\text{any}\;\zeta(1).
\end{align}

Following the F12 approach, the trial functions are expanded in a basis set comprising products of virtual orbitals (the standard part) and functions that explicitly depend on interelectron distance $r_{12}$ in terms of a correlation factor $f_{12}\equiv f_{12}(r_{12})$ (the F12 part). In the orbital invariant form of Klopper\cite{Klopper:91} they are expressed as
\begin{equation}
\label{eq:ansatz}
|\tilde{\tau}^{(ij)}\rangle
=\sum_{ab}c^{(ij)}_{ab}|\varphi^A_a\varphi^B_b\rangle
+\sum_{kl}t^{(ij)}_{kl}Q_{AB}f_{12}|\varphi^A_k\varphi^B_l\rangle,
\end{equation}
with unknown amplitudes $c^{(ij)}_{ab}$ and $t^{(ij)}_{kl}$. The standard part fulfills, by construction, the SO conditions, Eqs.~\eqref{eq:soA}-\eqref{eq:soB}, while in the F12 part the strong orthogonality is forced by the presence of the SO operator $Q_{AB}$, which in the \textit{Ansatz}~\textbf{3} of Ref.~\cite{Werner:07} reads
\begin{equation}
\label{eq:soproj}
Q_{AB}=(1-O)_A(1-O)_B(1-V_A V_B).
\end{equation}
This choice of the SO projector makes the standard and explicitly-correlated parts of the pair function mutually orthogonal. For further derivations it is convenient to represent $Q_{AB}$ in two formally equivalent forms:
\begin{equation}
\label{eq:soproj1}
Q^{(1)}_{AB}=1-P_{AB}
\end{equation}
with
\begin{equation}
\label{eq:pproj}
P_{AB}=O_A(1-P)_B+(1-P)_A O_B+P_AP_B,
\end{equation}
and
\begin{equation}
\label{eq:soproj2}
Q^{(2)}_{AB}=V_A(1-P)_B+(1-P)_A V_B+(1-P)_A(1-P)_B.
\end{equation}

\subsection{Amplitude equations}

Inserting Eq.~\eqref{eq:ansatz} into Eq.~\eqref{eq:functional} and differentiating with respect to $c^{(ij)}_{ab}$ and $t^{(ij)}_{kl}$, one obtains a coupled system of equations for amplitudes (Einstein convention for summation of repeated indices is assumed everywhere)
\begin{align}
\label{eq:ampc}
0&=G_{ab}^{ij}+(\varepsilon_a+\varepsilon_b-\varepsilon_i-\varepsilon_j)c^{(ij)}_{ab}+C_{ab}^{kl}\,t^{(ij)}_{kl},
\\
\label{eq:ampt}
0&=V_{kl}^{ij}+\left(B_{kl}^{mn}-(\varepsilon_i+\varepsilon_j)X_{kl}^{mn}\right)t^{(ij)}_{mn}+\left(C^T\right)_{kl}^{ab}\,c^{(ij)}_{ab},
\end{align}
where
\begin{align}
\label{eq:interG}
G_{ab}^{ij}&=\langle\varphi^A_a\varphi^B_b|r_{12}^{-1}|\varphi^A_i\varphi^B_j\rangle,\\
\label{eq:interV}
V_{kl}^{ij}&=\langle\varphi^A_k\varphi^B_l|f_{12}Q_{AB}r_{12}^{-1}|\varphi^A_i\varphi^B_j\rangle,\\
\label{eq:interX}
X_{kl}^{mn}&=\langle\varphi^A_k\varphi^B_l|f_{12}Q_{AB}f_{12}|\varphi^A_m\varphi^B_n\rangle,\\
\label{eq:interC}
C_{ab}^{kl}&=\langle\varphi^A_a\varphi^B_b|F_{AB}Q_{AB}f_{12}|\varphi^A_k\varphi^B_l\rangle,\\
\label{eq:interB}
B_{kl}^{mn}&=\langle\varphi^A_k\varphi^B_l|f_{12}Q_{AB}F_{AB}Q_{AB}f_{12}|\varphi^A_m\varphi^B_n\rangle.
\end{align}
Solving Eq.~\eqref{eq:ampc} for $c^{(ij)}_{ab}$, then solving Eq.~\eqref{eq:ampt} for $t^{(ij)}_{kl}$ produces full set of amplitudes which inserted in Eq.~\eqref{eq:paire} give closed form expression for the approximate pair energies\cite{Werner:07,Kong:12} 
\begin{equation}
\varepsilon_{ij}=
-4\sum_{ab}\frac{\left|G_{ab}^{ij}\right|^2}{\varepsilon_a+\varepsilon_b-\varepsilon_i-\varepsilon_j}
-4\left(\overline{V}^{(ij)}\right)^T\left(\overline{B}^{(ij)}\right)^{-1}\overline{V}^{(ij)},
\end{equation}
where matrix $\overline{B}^{(ij)}$ is defined by
\begin{equation}
\left(\overline{B}^{(ij)}\right)_{kl}^{mn}=B_{kl}^{mn}-(\varepsilon_i+\varepsilon_j)X_{kl}^{mn}
-\sum_{ab}\frac{\left(C^T\right)_{kl}^{ab}C_{ab}^{mn}}{\varepsilon_a+\varepsilon_b-\varepsilon_i-\varepsilon_j},
\end{equation}
and elements of vector $\overline{V}^{(ij)}$ have the form
\begin{equation}
\left(\overline{V}^{(ij)}\right)_{kl}=V_{kl}^{ij}-\sum_{ab}\frac{\left(C^T\right)_{kl}^{ab}G_{ab}^{ij}}{\varepsilon_a+\varepsilon_b-\varepsilon_i-\varepsilon_j}.
\end{equation}
Thus, finding the dispersion energy reduces to the problem of calculating the intermediates $G$, $V$, $X$, $C$, and $B$.

\subsection{\label{sec:GVX}Intermediates G, V, X}

The intermediate $G$, Eq.~\eqref{eq:interG}, is expressed in terms of the standard two-electron electron repulsion integrals. In contrast, exact calculation of intermediates $V$ and $X$, Eq.~\eqref{eq:interV} and \eqref{eq:interX}, respectively, would require three-electron integrals. Construction of these intermediates can be simplified by employing only the RI approximation. To this end, we use the SO projector in the $Q^{(1)}$ form and replace the exact projector $(1-P)_A$ in Eq.~\eqref{eq:pproj} by its approximate representation from Eq.~\eqref{eq:projX}  (analogous replacement is performed for the $(1-P)_B$ operator). 

The three intermediates can now be written in the explicit form involving only two-electron integrals and products of two-electron integrals
\begin{align}
G^{ij}_{ab}&=(g)^{ij}_{ab}\,,\\
V^{ij}_{kl}&=(fg)^{ij}_{kl}-(f)^{xb'}_{kl}(g)^{ij}_{xb'}-(f)^{a'y}_{kl}(g)^{ij}_{a'y}-(f)^{pq}_{kl}(g)^{ij}_{pq}\,,\\
X^{mn}_{kl}&=(f^2)^{mn}_{kl}-(f)^{xb'}_{kl}(f)^{mn}_{xb'}-(f)^{a'y}_{kl}(f)^{mn}_{a'y}-(f)^{pq}_{kl}(f)^{mn}_{pq}.
\end{align}
Here and in the following, we denote two-electron integrals with the integral kernel $k$ by
\begin{equation}
(k)^{\gamma\delta}_{\alpha\beta}=\langle\varphi^A_\alpha\varphi^B_\beta|k|\varphi^A_\gamma\varphi^B_\delta\rangle,
\end{equation}
and use the simplified notation: $f\equiv f_{12}$ and $g\equiv r_{12}^{-1}$.

\subsection{Variant B of the F12 approach}

The intermediates $C$ and $B$, Eq.~\eqref{eq:interC} and \eqref{eq:interB}, respectively, contain both the correlation factor and Fock operators. Exact calculation of these intermediates would require four-electron integrals. They are commonly treated with a combination of the commutator approach\cite{Kutzelnigg:85,Kutzelnigg:91} and the RI approximation. In the commutator approach we start with splitting the general Fock operator for a closed-shell system in two parts: $F=(F+K)-K$, where $K$ is the nonlocal exchange operator. The first part includes only the kinetic energy operator $T$ and two local operators: electron-nuclei attraction and Coulomb operator. Therefore, the product of $(F+K)_{AB}$ with the correlation factor can be rewritten as
\begin{equation}
\label{eq:commut}
(F+K)_{AB}f_{12}=[T_{12},f_{12}]+f_{12}(F+K)_{AB}.
\end{equation}
The integrals with the two-electron kernel $[T_{12},f_{12}]$ (denoted further as $tf$ in the simplified notation for integral kernels) can be calculated analytically.

Inserting the $Q^{(1)}_{AB}$ form of the SO projector into Eq.~\eqref{eq:interC} and using the commutator approach in the $F_{AB}f_{12}$ term, we obtain
\begin{equation}
C_{ab}^{kl}=\langle\varphi^A_a\varphi^B_b|
[T_{12},f_{12}]+f_{12}(F+K)_{AB}-K_{AB}f_{12}-F_{AB}P_{AB}f_{12}
|\varphi^A_k\varphi^B_l\rangle
\end{equation}
By applying the approximate RI operator, Eq.~\eqref{eq:proj1}, the intermediate $C$ can now be expressed in terms of two-electron integrals and products of two-electron integrals and matrix elements of the exchange and Fock operators
\begin{equation}
\label{eq:CinB}
\begin{split}
C^{kl}_{ab}&=(tf)^{kl}_{ab}
+(f)^{\tilde kl}_{ab}+(f)^{k\tilde l}_{ab} \\
&-(K_A)^{p'}_a(f)^{kl}_{p'b}-(f)^{kl}_{aq'}(K_B)^{q'}_b
-(F_A)^c_a(f)^{kl}_{cb}-(f)^{kl}_{ad}(F_B)^d_b.
\end{split}
\end{equation}
The indices with tildes represent orbitals from the AMO space that were transformed according to
\begin{equation}
\varphi^A_{\tilde k}=\sum_{p'}\left((F+K)_A\right)^k_{p'}\varphi^A_{p'},\quad
\varphi^B_{\tilde l}=\sum_{q'}\left((F+K)_B\right)^l_{q'}\varphi^B_{q'}.
\end{equation}
In deriving last two terms in Eq.~\eqref{eq:CinB}, the Brillouin condition was employed: \mbox{$OFV=VFO=0$}.

To evaluate the intermediate $B$, we split the Fock operators and represent $(F+K)_{AB}Q_{AB}f_{12}$ by
\begin{equation}
(F+K)_{AB}Q_{AB}f_{12}=Q_{AB}[T_{12},f_{12}]+Q_{AB}f_{12}(F+K)_{AB}+[(F+K)_{AB},Q_{AB}]f_{12}
\end{equation}
obtaining
\begin{equation}
\label{eq:B1}
\begin{split}
B_{kl}^{mn}
&=\frac12\,\langle\varphi^A_k\varphi^B_l|f_{12}Q_{AB}[T_{12},f_{12}]+[f_{12},T_{12}]Q_{AB}f_{12}|\varphi^A_m\varphi^B_n\rangle\\
&+\frac12\,\langle\varphi^A_k\varphi^B_l|f_{12}Q_{AB}f_{12}(F+K)_{AB}+(F+K)_{AB}f_{12}Q_{AB}f_{12}|\varphi^A_m\varphi^B_n\rangle\\
&-\frac12\,\langle\varphi^A_k\varphi^B_l|f_{12}Q_{AB}(F+K)_{AB}P_{AB}f_{12}+f_{12}P_{AB}(F+K)_{AB}Q_{AB}f_{12}|\varphi^A_m\varphi^B_n\rangle\\
&-\langle\varphi^A_k\varphi^B_l|f_{12}Q_{AB}K_{AB}Q_{AB}f_{12}|\varphi^A_m\varphi^B_n\rangle,
\end{split}
\end{equation}
where the identity $Q_{AB}[(F+K)_{AB},Q_{AB}]=-Q_{AB}(F+K)_{AB}P_{AB}$ was used to simplify the third term. The first two terms in Eq.~\eqref{eq:B1} can be treated similarly as in Sec.~\ref{sec:GVX}, but with the last two terms we proceed further using the SO projector in the $Q^{(2)}_{AB}$ form. After rearrangement they read
\begin{equation}
\label{eq:B2}
\begin{split}
-\frac12\,\langle\varphi^A_k\varphi^B_l|f_{12}\Big[&\big((1-P)K+K(1-P)\big)_AV_B \\
+&\big((1-P)K+K(1-P)+VK+KV\big)_A(1-P)_B\\
+&\big(A\leftrightarrow B\big)\Big]f_{12}|\varphi^A_m\varphi^B_n\rangle \\
-\frac12\,\langle\varphi^A_k\varphi^B_l|f_{12}\Big[&\big((1-P)FO+OF(1-P)\big)_AV_B \\
+&\big((1-P)FO+OF(1-P)+VFO+OFV\big)_A(1-P)_B\\
+&\big(A\leftrightarrow B\big)\Big]f_{12}
|\varphi^A_m\varphi^B_n\rangle\\
-\frac12\,\langle\varphi^A_k\varphi^B_l|f_{12}&Q_{AB}F_{AB}V_AV_Bf_{12}+f_{12}V_AV_BF_{AB}Q_{AB}f_{12}|\varphi^A_m\varphi^B_n\rangle.
\end{split}
\end{equation}
The third term was separated as it replaces the EBC-like terms $(1-P)FV$ and $VF(1-P)$ that otherwise would appear in the second term of Eq.~\eqref{eq:B2} by an expression that can be calculated reusing matrix elements of the intermediate $C$.~\cite{Werner:07}

In the RI approximation the final expression for the intermediate $B$ in variant B is
\begin{equation}
\begin{split}
B^{mn}_{kl}&=(ftf)^{mn}_{kl}-\frac12\left((f)^{xb'}_{kl}(tf)^{mn}_{xb'}+(f)^{a'y}_{kl}(tf)^{mn}_{a'y}+(f)^{pq}_{kl}(tf)^{mn}_{pq} +\text{h.c.}\right) \\
&+\frac12\left((f^2)^{\tilde mn}_{kl}-(f)^{xb'}_{kl}(f)^{\tilde mn}_{xb'}-(f)^{a'y}_{kl}(f)^{\tilde mn}_{a'y}-(f)^{pq}_{kl}(f)^{\tilde mn}_{pq} +\text{h.c.}\right)\\
&+\frac12\left((f^2)^{m\tilde n}_{kl}-(f)^{xb'}_{kl}(f)^{m\tilde n}_{xb'}-(f)^{a'y}_{kl}(f)^{m\tilde n}_{a'y}-(f)^{pq}_{kl}(f)^{m\tilde n}_{pq} +\text{h.c.}\right)\\
&-\frac12\left((f)^{a'b}_{kl}(K_A)^{p'}_{a'}(f)^{mn}_{p'b}+(f)^{ab'}_{kl}(K_B)^{q'}_{b'}(f)^{mn}_{aq'}+\text{h.c.}\right)\\
&-\frac12\left((f)^{a'b'}_{kl}(K_A)^{p'}_{a'}(f)^{mn}_{p'b'}+(f)^{a'b'}_{kl}(K_B)^{q'}_{b'}(f)^{mn}_{a'q'}+\text{h.c.}\right)\\
&-\frac12\left((f)^{ab'}_{kl}(K_A)^{p'}_{a}(f)^{mn}_{p'b'}+(f)^{a'b}_{kl}(K_B)^{q'}_{b}(f)^{mn}_{a'q'}+\text{h.c.}\right)\\
&-\frac12\left((f)^{a'b}_{kl}(F_A)^{x}_{a'}(f)^{mn}_{xb}+(f)^{ab'}_{kl}(F_B)^{y}_{b'}(f)^{mn}_{ay}+\text{h.c.}\right)\\
&-\frac12\left((f)^{a'b'}_{kl}(F_A)^{x}_{a'}(f)^{mn}_{xb'}+(f)^{a'b'}_{kl}(F_B)^{y}_{b'}(f)^{mn}_{a'y}+\text{h.c.}\right)\\
&-\frac12\left((f)^{ab}_{kl}C^{mn}_{ab}+\text{h.c.}\right),
\end{split}
\end{equation}
where $ftf$ denotes two-electron integrals with the integral kernel $[f_{12},[T_1,f_{12}]]=[f_{12},[T_2,f_{12}]]$ and the Brillouin condition was employed.

\subsection{Variant A of the F12 approach}

In the present approach we adopt the definition of variant A as proposed in Ref.~\cite{Werner:07}. The intermediate $C$ is calculated in the same way as in variant B, Eq.~\eqref{eq:CinB}, but in the case of intermediate $B$ additional approximations are assumed. First, we entirely neglect the commutator of the exchange operator and correlation factor, that is we set $[K_{AB},f_{12}]=0$. Second, we assume that the generalized Brillouin condition is fulfilled by the occupied orbitals, that is $OF(1-P)=(1-P)FO=0$. In variant A, the intermediate $B$ is obtained from the formula
\begin{equation}
\begin{split}
B^{mn}_{kl}&=(ftf)^{mn}_{kl}-\frac12\left((f)^{xb'}_{kl}(tf)^{mn}_{xb'}+(f)^{a'y}_{kl}(tf)^{mn}_{a'y}+(f)^{pq}_{kl}(tf)^{mn}_{pq} +\text{h.c.}\right) \\
&+\frac12\left((f^2)^{\bar mn}_{kl}-(f)^{xb'}_{kl}(f)^{\bar mn}_{xb'}-(f)^{a'y}_{kl}(f)^{\bar mn}_{a'y}-(f)^{pq}_{kl}(f)^{\bar mn}_{pq} +\text{h.c.}\right)\\
&+\frac12\left((f^2)^{m\bar n}_{kl}-(f)^{xb'}_{kl}(f)^{m\bar n}_{xb'}-(f)^{a'y}_{kl}(f)^{m\bar n}_{a'y}-(f)^{pq}_{kl}(f)^{m\bar n}_{pq} +\text{h.c.}\right)\\
&-\frac12\left((f)^{ab}_{kl}C^{mn}_{ab}+\text{h.c.}\right),
\end{split}
\end{equation}
where the indices with bars represent orbitals from the MO space that were transformed according to
\begin{equation}
\varphi^A_{\bar m}=\sum_{p}(F_A)^m_{p}\varphi^A_{p},\quad
\varphi^B_{\bar n}=\sum_{q}(F_B)^n_{q}\varphi^B_{q}.
\end{equation}

\subsection{Variant C of the F12 approach}

The idea of variant C is to avoid entirely the calculation of integrals with a single commutator operator $[T_{12},f_{12}]$. This is accomplished by relying more heavily on the RI approximation in deriving working formulas for intermediates $C$ and $B$.~\cite{Kedzuch:05}

In the case of the intermediate $C$, it suffices to insert in Eq.~\eqref{eq:interC} the SO operator in the $Q^{(2)}_{AB}$ form and use the approximate projection on the complementary space, Eq.~\eqref{eq:projX}, to get
\begin{equation}
C^{kl}_{ab}=(F_A)^{a'}_a(f)^{kl}_{a'b}+(f)^{kl}_{ab'}(F_B)^{b'}_b.
\end{equation}

The derivation for the intermediate $B$ starts with inserting in Eq.~\eqref{eq:interB} the SO operator in the $Q^{(1)}_{AB}$ form
\begin{equation}
\begin{split}
B_{kl}^{mn}&=\langle\varphi^A_k\varphi^B_l|f_{12}F_{AB}f_{12}|\varphi^A_m\varphi^B_n\rangle\\
&-\langle\varphi^A_k\varphi^B_l|f_{12}P_{AB}F_{AB}f_{12}+f_{12}F_{AB}P_{AB}f_{12}-f_{12}P_{AB}F_{AB}P_{AB}f_{12}|\varphi^A_m\varphi^B_n\rangle.
\end{split}
\end{equation}
The first term can be calculated by splitting the Fock operator and expressing $f_{12}(F+K)_{AB}f_{12}$ through the double commutator expansion. The second term can be written in many equivalent ways.
The representation used by us
\begin{equation}
\label{eq:PinC}
\begin{split}
-\langle\varphi^A_k\varphi^B_l|f_{12}\Big[& \big(PFP+(1-P)FP+PF(1-P)\big)_A P_B\\
+&\big((1-P)F(1-P)\big)_A O_B\\
+&\big(OFO +OFV+VFO +(1-P)FO+OF(1-P)\big)_A(1-P)_B\\
+&\big(A\leftrightarrow B\big)\big)\Big]f_{12}|\varphi^A_m\varphi^B_n\rangle.
\end{split}
\end{equation}
 differs slightly from the one adopted in Ref.~\cite{Kedzuch:05}.

In the RI approximation the final expression for the intermediate $B$ in variant C is
\begin{equation}
\begin{split}
B^{mn}_{kl}&=(ftf)^{mn}_{kl}+\frac12\left((f^2)^{\tilde mn}_{kl}+(f^2)^{m\tilde n}_{kl}+\text{h.c.}\right)\\
&-\left((f)^{p'q'}_{kl}(K_A)^{r'}_{p'}(f)^{mn}_{r'q'}+(f)^{p'q'}_{kl}(K_B)^{s'}_{q'}(f)^{mn}_{p's'}\right)\\
&-\left((f)^{pq}_{kl}(F_A)^{r}_{p}(f)^{mn}_{rq}+(f)^{pq}_{kl}(F_B)^{s}_{q}(f)^{mn}_{ps}\right)\\
&-\left((f)^{a'q}_{kl}(F_A)^{r}_{a'}(f)^{mn}_{rq}+(f)^{pb'}_{kl}(F_B)^{s}_{b'}(f)^{mn}_{ps}+\text{h.c.}\right)\\
&-\left((f)^{a'y}_{kl}(F_A)^{c'}_{a'}(f)^{mn}_{c'y}+(f)^{xb'}_{kl}(F_B)^{d'}_{b'}(f)^{mn}_{xd'}\right)\\
&-\left((f)^{xb'}_{kl}(F_A)^{z}_{x}(f)^{mn}_{zb'}+(f)^{a'y}_{kl}(F_B)^{w}_{y}(f)^{mn}_{a'w}\right)\\
&-\left((f)^{a'b'}_{kl}(F_A)^{z}_{a'}(f)^{mn}_{zb'}+(f)^{a'b'}_{kl}(F_B)^{w}_{b'}(f)^{mn}_{a'w}+\text{h.c.}\right),
\end{split}
\end{equation}
where the Brillouin condition was employed to remove the $OFV$ and $VFO$ terms from Eq.~\eqref{eq:PinC}.

\section{\label{sec:details}Technical details}

\subsection{Computer implementation}

Equations of the $E^{(20)}_\mathrm{disp}$(F12) theory in variants A, B, and C were implemented in the in-house Fortran 2008 code. In calculations we used three forms of the correlation factor: a long-range linear one $r_{12}$ (abbreviated R12), and two types of short-range factors $\exp(-\gamma r_{12})$ (STG), and $r_{12}\exp(-\gamma r_{12})$ (RSTG). For the exponential factors we did not employ their expansion in a set of Gaussian functions. The integrals with scalar kernels $f$, $fg$, $f^2$, $ftf\equiv (\nabla_1 f_{12})^2$ and Gaussian basis functions can all be calculated using a unified recursion scheme provided that the primitive integral expressions for $s$ functions are known.~\cite{Ahlrichs:06,Silkowski:15} The formulas for the primitive integrals in the case of the $r_{12}$ factor were taken from Ref.~\cite{Ahlrichs:06}, and in the case of the exponential factors from Ref.~\cite{Ten-no:04,Ten-no:07}. We used the implementation of the Obara-Saika recursions\cite{Obara:86,Obara:88} available in the \textsc{LibInt} library.~\cite{Libint2} The integrals with the $tf$ kernel were constructed from second derivatives of integrals with the $f$ kernel.~\cite{Samson:02} To store and organize data on disc the \textsc{HDF5} library was used.~\cite{HDF5}

\subsection{Computational details}

In order to assess the accuracy of the  $E^{(20)}_\mathrm{disp}$(F12) method we performed calculations for systems from the A24 data set\cite{Rezac:13} comprising 24 small dimers. They represent three classes of noncovalent interactions: hydrogen bonded dimers (six complexes collectively referred to as class-H), mixed electrostatic/dispersion dimers (ten complexes, class-M), and dispersion-dominated dimers (nine complexes, class-D). Complexes in the A24 data set are given in near equilibrium geometries. By $R_\mathrm{eq}$ we denote the distance between the center-of-masses of the interacting monomers.

All explicitly correlated calculations employed aug-cc-pV$X$Z basis sets ($X$ = D, T) of Dunning and co-workers~\cite{Dunning:89} together with matching complementary auxiliary basis sets aug-cc-pV$X$Z\_OPTRI of Yousaf and Peterson.~\cite{Yousaf:09} Good performance of these basis sets was recently reported by Sirianni \textit{et al.}\cite{Sirianni:17} who analyzed the accuracy of supermolecular noncovalent interaction energies obtained with CCSD-F12(T**) methods (T** denoting the scaled-triples correction).

Reference values for $E^{(20)}_\mathrm{disp}$ were obtained by performing CBS extrapolation of the orbital-based results from the aug-cc-pVQZ and aug-cc-pV5Z basis sets according to the two-point scheme of Halkier et al.~\cite{Halkier:98}
\begin{equation}
\label{eq:CBS}
E^{(20)}_{\rm disp, CBS} = 
\frac{X^3\, E^{(20)}_{{\rm disp},X} - (X-1)^3\, E^{(20)}_{{\rm disp},X-1} }
{X^3-(X-1)^3}.
\end{equation}

The error statistics is performed in terms of relative percent errors defined as
\begin{equation}
\label{eq:error}
\Delta=\frac{E^{(20)}_{\mathrm{disp},X}\mathrm{(F12)}-E^{(20)}_\mathrm{disp,\mathrm{CBS}}}{|E^{(20)}_\mathrm{disp,\mathrm{CBS}}|}\cdot 100\%.
\end{equation}
Because the dispersion energies for complexes in the A24 data set vary significantly (from $-3.5$~kJ/mol for the water-methane system to $-21.0$~kJ/mol for the formaldehyde dimer) we avoid error measures in absolute units which would bias the analysis towards systems with large $E^{(20)}_\mathrm{disp}$.

A detailed analysis of the basis set convergence of $E^{(20)}_\mathrm{disp}$(F12) was performed only for the water dimer. In this case basis sets with cardinal number $X$ up to $X = \mathrm{Q}$ were employed. To obtain more strict estimation of the reference dispersion energy we adopted the extrapolation from basis sets aug-cc-pV5Z and aug-cc-pV6Z using Eq.~\eqref{eq:CBS}.

\section{\label{sec:results}Results and discussion}

In this section we analyze $\gamma$-dependence of the explicitly correlated dispersion energy for the exponential correlation factors and make recommendations on the optimal value of the length-scale parameter $\gamma$. Then, we examine the performance of all combinations of correlation factors and A/B/C variants of the F12 approach, and identify the best-performing $E^{(20)}_\mathrm{disp}$(F12) scheme. Basis set convergence is discussed in the case of the water dimer. Finally, we analyze the applicability of the method for large intermolecular distances, that is $1.5\,R_\mathrm{eq}$ and $2.0\,R_\mathrm{eq}$.

\subsection{Dependence on the length-scale parameter}

The short-range exponential factors STG and RSTG depend on a single length-scale parameter $\gamma$. Due to the variational character of the second-order dispersion energy, the optimal value of $\gamma$ may be fixed as the value corresponding to the minimum of $E^{(20)}_\mathrm{disp}$(F12). 

\begin{figure*}
\caption{\label{fig:skan} Dependence of the calculated $E^{(20)}_{\mathrm{disp}}$(F12) dispersion energy on the value of the length-scale parameter $\gamma$ for the water dimer (geometry from the A24 data set). Solid and dashed lines represent values obtained with the aug-cc-pVDZ and aug-cc-pVTZ basis sets, respectively. Horizontal line marks the $E^{(20)}_\mathrm{disp,CBS}$ reference.} 
\includegraphics[width=\textwidth]{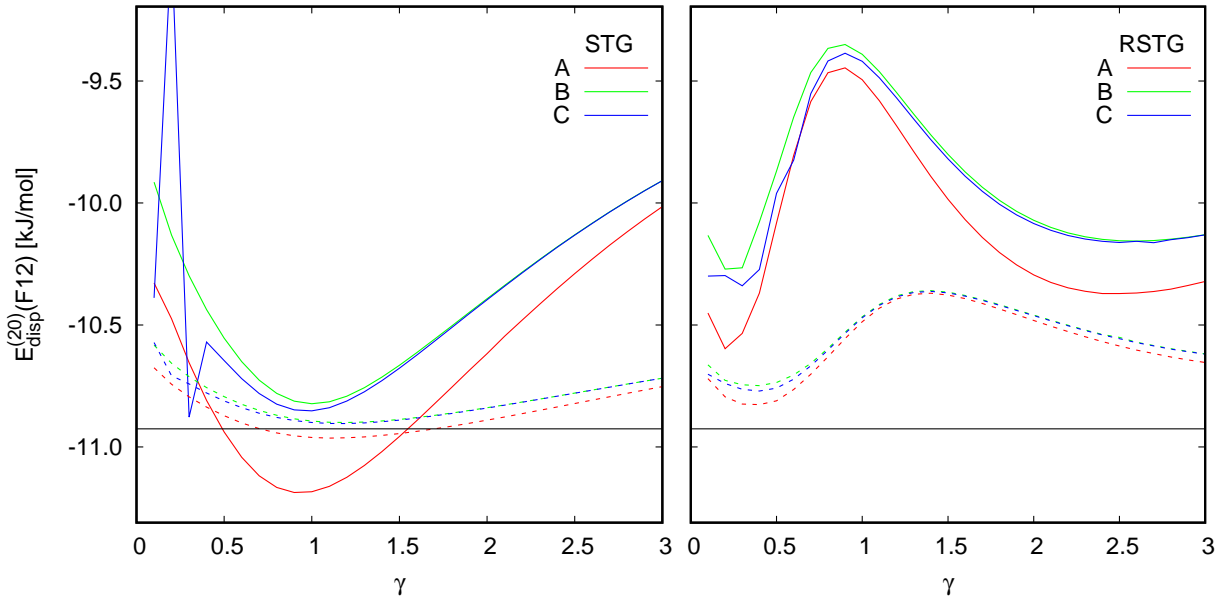}
\end{figure*}

We investigated $\gamma$-dependence of the dispersion energy for all complexes from the A24 data set using the aug-cc-pVDZ basis set. Additionally, with aug-cc-pVTZ we performed calculations for the water dimer (representative of the class-H), water-methane system (class-M), and methane dimer (class-D). A grid of 30 $\gamma$ values ranging from 0.1 to 3.0 with 0.1 interval was assumed in all cases. The results for the water dimer, which illustrate general trends for molecules in the A24 data set, are shown in Figure~\ref{fig:skan}.

Because of the approximations adopted in variant A of the F12 approach the dispersion energy obtained in this case lacks variational character for some values of $\gamma$ when the correlation factor is of the STG form. In Figure~\ref{fig:skan} this is manifested as $E^{(20)}_\mathrm{disp}$(F12) laying below the reference for $\gamma$ in the $0.5\!-\!1.5$ range. A similar phenomenon is less pronounced for the RSTG factor and occurs mainly for the class-D complexes. Therefore, variant A is excluded from further search for the optimal $\gamma$ value. 

The results calculated in variants B and C closely follow each other, especially for $\gamma>1.0$. With the STG correlation factor the difference between B and C energies slowly increases with decreasing $\gamma$ for $\gamma<1.0$. At $\gamma=0.1$ and with the aug-cc-pVDZ basis set the difference does not exced 5\% of the reference dispersion energy and with aug-cc-pVTZ it reduces to 0.5\%. The differences between B and C variants with the RSTG correlation factor are smaller than with the STG factor when the aug-cc-pVDZ basis set is used, and similar with aug-cc-pVTZ. Energies in variant C obtained with the aug-cc-pVDZ basis set exhibit irregular behavior, similar to that observed in Figure~\ref{fig:skan}, in half of the analysed complexes. In contrast, variant B with this basis set and both B and C variants in the aug-cc-pVTZ basis set provide dispersion energies which are smooth functions of $\gamma$. Significant irregularities are observed only for dimers containing the argon atom. 

Let us examine the $\gamma$-dependence for the STG correlation factor in more detail. With the aug-cc-pVDZ basis set the optimal values of $\gamma$
are: $\gamma\approx 1.0$ (class-H complexes), $\gamma\approx 0.8-0.9$ (class-M), and $\gamma\approx 0.8$ (class-D). With the aug-cc-pVTZ basis set, the optimal $\gamma$ shifts to larger values: from 1.0 to 1.2 for the water dimer, from 0.9 to 1.0 for the water-methane dimer, and from 0.8 to 0.9 for the methane dimer. We recommend using a single short-range parameter $\gamma=0.9$ for the STG correlation factor. This choice is optimal for the largest set of complexes in the aug-cc-pVDZ basis set. Because of generally weaker $\gamma$-dependence of the dispersion energy in aug-cc-pVTZ, this choice provides near-optimal results also in this case.

When the correlation factor has the RSTG form, the plot of the calculated dispersion energy as a function of $\gamma$ exhibits a double-well character. For class-H dimers the second (large-$\gamma$) minimum is shallower than the first (small-$\gamma$) one. For other classes both minima are of similar depth. The location of the second minimum depends strongly on both the system ($\gamma\approx 2.7$ for the water-ammonia dimer compared to $\gamma\approx 1.5$ for the methane dimer) and basis set (the second minimum for the water dimer is at $\gamma\approx 2.6$ with the aug-cc-pVDZ basis set and beyond $\gamma=3.0$ with aug-cc-pVTZ). Therefore, the large-$\gamma$ minimum is unsuitable for establishing a universal optimal value of $\gamma$ for the RSTG correlation factor. In contrast, the position of the small-$\gamma$ minimum is stable. With the aug-cc-pVDZ basis set it occurs around $\gamma=0.2$ for all considered systems. With the aug-cc-pVTZ basis set the optimal $\gamma$ shifts to larger values only for the water dimer ($\gamma\approx 0.4$) and the water-methane dimer ($\gamma\approx 0.3$). Again, due to the weaker $\gamma$-dependence of the dispersion energy in the aug-cc-pVTZ basis set, we recommend adopting a single universal value $\gamma=0.2$.

All calculations in the following sections were performed with the recommended values of the length-scale parameters, that is $\gamma=0.9$ for the STG correlation factor, and $\gamma=0.2$ for the RSTG correlation factor.

\subsection{\label{sec:equil}F12 dispersion energy for the A24 data set}

\begin{figure}
\caption{\label{fig:gaus} Normal distributions of the errors in the $E^{(20)}_\mathrm{disp}$(F12) dispersion energy for molecules from the A24 data set obtained using various combinations of (correlation factor)/(F12 variant).
}
\includegraphics[width=\columnwidth]{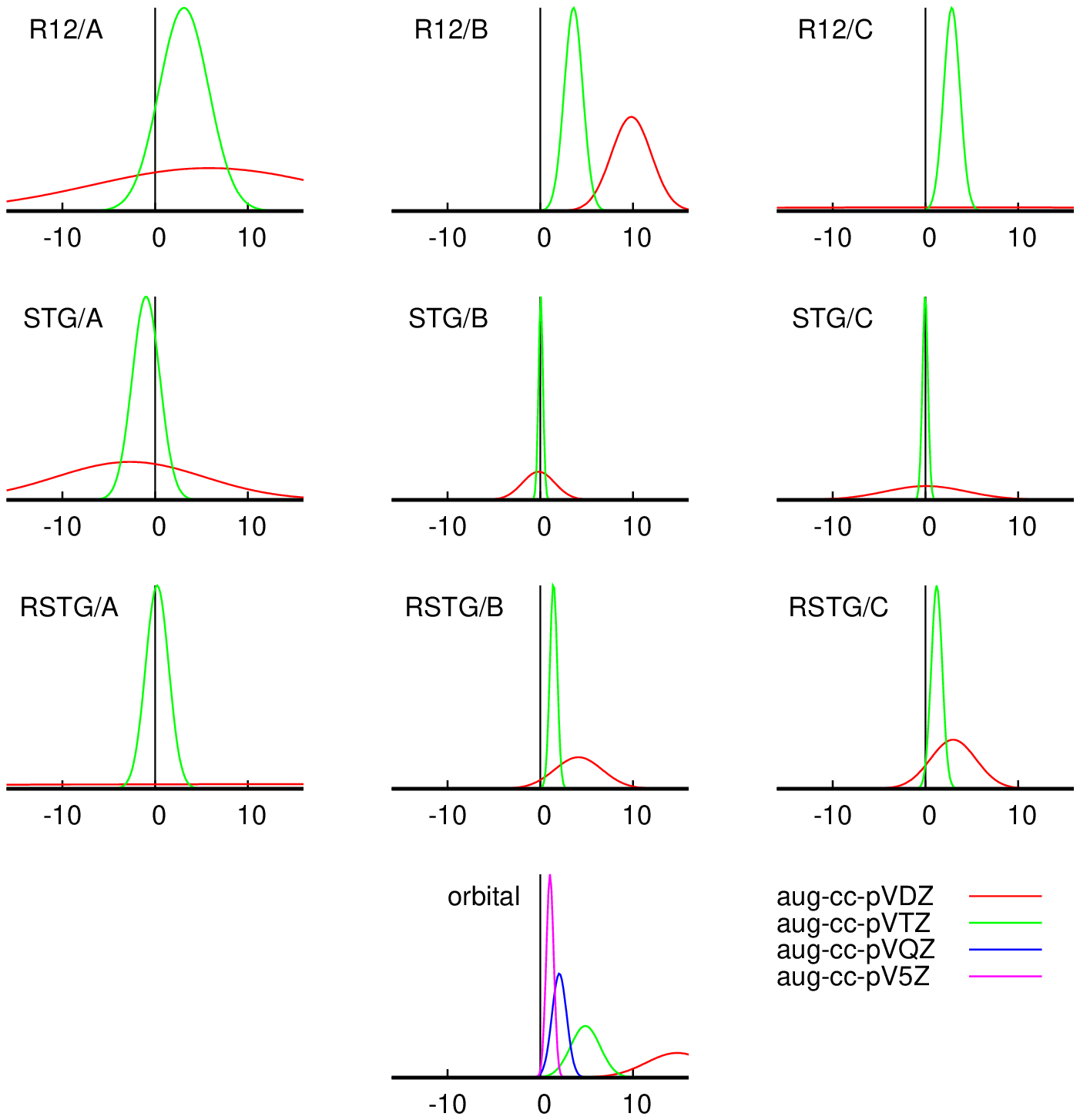}
\end{figure}

In Table~\ref{tab:stat} we present the error statistics for the A24 data set in terms of the mean error $\overline{\Delta}$, the standard deviation $\sigma$, the mean absolute error $\overline{\Delta}_\mathrm{abs}$, and the maximum absolute error $\Delta_\mathrm{max}$. According to the definition of the relative percent error given in Eq.~\eqref{eq:error} and attractive character of the dispersion energy, positive and negative values of $\overline{\Delta}$ correspond to under- and overestimation of the magnitude of $E^{(20)}_\mathrm{disp}$, respectively. In Figure~\ref{fig:gaus}, the mean error and standard deviation are depicted assuming normal distribution of errors. We show results for all possible combinations of correlation factors and F12 variants considered in this work. Additionally, the error statistics for $E^{(20)}_\mathrm{disp}$ calculated using the standard orbital approach is shown for comparison.

\begin{table}
\caption{\label{tab:stat} Summary error statistics (in percent) for the calculated $E^{(20)}_\mathrm{disp}$(F12) dispersion energy for dimers from the A24 data set obtained using various combinations of (correlation factor)/(F12 variant). $X = \mathrm{D, T}$ denote aug-cc-pVDZ and aug-cc-pVTZ basis sets, respectively.}
\begin{tabular}{ccrrrr}
$X$ & A/B/C & $\overline{\Delta}\;\;$ & $\sigma\;\;$ & $\overline{\Delta}_\mathrm{abs}$ & $\Delta_\mathrm{max}$ \\
\hline
\\[-1ex]
&\multicolumn{4}{l}{R12} \\
\\[-1ex]
D &A&   5.83 &  12.62 &  10.45 &  40.32 \\
T &A&   3.11 &   2.67 &   3.17 &  15.02 \\
\\[-1ex]
D &B&   9.82 &   2.15 &   9.82 &  15.71 \\
T &B&   3.59 &   1.00 &   3.59 &   6.16 \\
\\[-1ex]
D &C&   2.50 &  48.86 &  20.97 & 209.69 \\
T &C&   2.85 &   0.87 &   2.85 &   4.25 \\
\\[-1ex]
&\multicolumn{4}{l}{STG($\gamma=0.9$)} \\
\\[-1ex]
D &A&  $-$2.71 &   7.98 &   5.29 &  31.07 \\
T &A&  $-$0.97 &   1.49 &   0.97 &   7.25 \\
\\[-1ex]
D &B&  $-$0.17 &   1.74 &   1.10 &   5.30 \\
T &B&   0.06 &   0.24 &   0.18 &   0.68 \\
\\[-1ex]
D &C&   0.06 &   4.42 &   2.01 &  18.52 \\
T &C&  $-$0.02 &   0.29 &   0.19 &   0.93 \\
\\[-1ex]
&\multicolumn{4}{l}{RSTG($\gamma=0.2$)} \\
\\[-1ex]
D &A&  11.76 &  56.12 &  13.64 & 274.81 \\
T &A&   0.23 &   1.25 &   0.92 &   4.43 \\
\\[-1ex]
D &B&   4.11 &   2.57 &   4.23 &  11.16 \\
T &B&   1.43 &   0.39 &   1.43 &   2.48 \\
\\[-1ex]
D &C&   2.99 &   2.46 &   3.41 &   7.53 \\
T &C&   1.22 &   0.59 &   1.22 &   3.44 \\
\\[-1ex]
&\multicolumn{4}{l}{orbital approach} \\
\\[-1ex]
D &&  14.78 &   3.34 &  14.78 &  21.68 \\
T &&   4.86 &   1.59 &   4.86 &   8.99 \\
Q &&   2.06 &   0.79 &   2.06 &   4.49 \\
5 &&   1.05 &   0.40 &   1.05 &   2.30 \\
\end{tabular}
\end{table}

The magnitude of the dispersion energy calculated with the R12 correlation factor is systematically underestimated ($\overline{\Delta}$ is positive regardless of the basis set and F12 variant). Increasing the basis set size reduces the mean error in variants A and B by a factor of 2 and 3, respectively, while in variant C we observe a slight increase in $\overline{\Delta}$. The standard deviation is always smaller in the larger basis set. The reduction in the $\sigma$ value for variant C is two orders of magnitude, one order of magnitude for variant A, but only by a factor of two for variant B. The statistics for variant C in aug-cc-pVDZ is distorted by a single outlier (the dispersion energy for the methane-ethane dimer is twice as large as the reference value). After removing this point both $\overline{\Delta}=11.72$ and $\sigma=18.98$ are closer to the results for variants A and B. With the aug-cc-pVTZ basis set the performance of all three variants of F12 theory is similar in terms of the mean error, but variants B and C are preferred due to the smaller standard deviation. Nevertheless, the choice of the R12 correlation factor offers only marginal improvement over the standard approach in the same basis set.

The STG factor, the excellent performance of which is observed for intramolecular correlation, proves to be a preferred choice also for intermolecular correlation. The absolute value of the mean errors remains below 1\% for nearly all combinations of the correlation factor and F12 variant, the only exception being variant A in the aug-cc-pVDZ basis set. Although variant A systematically overestimates the magnitude of the dispersion energy, it still offers a substantial improvement over standard orbital calculations---$|\overline{\Delta}|$ for a basis set with a given cardinal number $X$ is comparable to results from the orbital calculations with $(X\!+2)$ basis. Variants B and C reproduce reference energies with a remarkable accuracy---in both basis sets we obtain $|\overline{\Delta}|<0.2\%$. Moreover, with the aug-cc-pVTZ basis set the standard deviation stays below 0.3\%, and the mean absolute errors and maximal absolute errors are below 0.2\% and 1.0\%, respectively. This indicates that the description of the dispersion energy is consistent for the entire A24 data set. It should  be stressed that the explicitly correlated dispersion energy obtained with variants B and C surpasses the accuracy of the orbital calculations in aug-cc-pV5Z. The superior performance of the STG/B and STG/C approaches, particularly in the aug-cc-pVTZ, is evidenced in Figure~\ref{fig:gaus} by sharp distributions located at the origin. 

The RSTG correlation factor can be understood as a generalization of the R12 form. The latter is a limiting case of the RSTG factor for $\gamma\rightarrow 0$. The overall underestimation of the magnitude of the dispersion energy, observed with the R12 factor, can also be recognized for RSTG. However, choosing the optimal value of the length-scale parameter $\gamma$ leads to better error statistics. This trend is not clear for variant A in aug-cc-pVDZ unless a single outlier is removed. When the HF-methane dimer is excluded from the analysis, the mean error reduces to 0.32\% and standard deviation to 3.18\%. Then, the values of $\overline{\Delta}$ calculated in variant A with aug-cc-pVDZ and aug-cc-pVTZ become similar to each other and close to zero. In variants B and C the mean error and the standard deviation diminish with the increasing basis set size. In the aug-cc-pVTZ basis set both $\overline{\Delta}$ and $\sigma$ are approximately two times smaller than in the case of the R12 factor. Nevertheless, the RSTG factor does not provide accuracy levels that can be reached with the STG factor. The mean errors of $E^{(20)}_\mathrm{disp}$(F12) obtained with RSTG are two orders of magnitude larger than the values observed with STG. The standard deviations are also larger with the RSTG factor but only by a factor of two. The overall accuracy of $E^{(20)}_\mathrm{disp}$(F12) with RSTG calculated in basis set with cardinal number $X$ falls between the accuracy of the orbital approach in basis sets with $(X\!+1)$ and $(X\!+2)$.

The complexes from the A24 data set are grouped into three classes according to the dominant character of the interaction. It is interesting to examine the accuracy of the dispersion energy calculated in the explicitly correlated approach for complexes of each class separately. The results for the best-performing combinations of the correlation factor and F12 approach, that is STG/B and STG/C, are presented in Table~\ref{tab:stat2}.

\begin{table}
\caption{\label{tab:stat2} Relative percent errors in the calculated dispersion
energy $E^{(20)}_\mathrm{disp}$(F12) for dimers from the A24 data set grouped according to the dominant character of the intermolecular interaction. $X = \mathrm{D, T}$ denote aug-cc-pVDZ and aug-cc-pVTZ basis sets, respectively.}
\begin{tabular}{ccrrrrrr}
&&
\multicolumn{2}{c}{class-H}&
\multicolumn{2}{c}{class-M}&
\multicolumn{2}{c}{class-D}\\[0.5ex]
$X$ & method & 
$\overline{\Delta}\;\;$ & $\sigma\;\;$ &
$\overline{\Delta}\;\;$ & $\sigma\;\;$ &
$\overline{\Delta}\;\;$ & $\sigma\;\;$ \\
\hline
\\[-1ex]
D&STG/B& 0.94 & 0.69 & 0.44 & 0.75 & $-$1.47 & 2.17 \\
T&STG/B& 0.33 & 0.21 & 0.07 & 0.08 & $-$0.11 & 0.23 \\
\\[-1ex]
D&STG/C& 0.11 & 0.77 & $-$0.68 & 2.78 & 0.85 & 6.76 \\
T&STG/C& 0.28 & 0.20 & $-$0.01 & 0.18 & $-$0.21 & 0.31 \\ 
\end{tabular}
\end{table}

For hydrogen bonded dimers, the mean errors and standard deviations are consistently smaller than 1\%. The accuracy of both F12 approaches is similar, with the exception of $\overline{\Delta}$ in the aug-cc-pVDZ basis set. The magnitude of the dispersion energy in aug-cc-pVTZ is slightly underestimated (ca.~0.3\%). This is most likely related to the fact that the value $\gamma=0.9$ recommended for the entire A24 data set is suboptimal for class-H dimers. For example, in the case of the water dimer replacing the recommended $\gamma$ with the optimal one ($\gamma=1.2$) reduces the relative error from 0.37\% to 0.24\% for STG/B and from 0.31\% to 0.20\% for STG/C.

The performance of the STG/B approach for complexes of class-M (mixed electrostatic/dispersion interactions) is excellent. In the aug-cc-pVDZ basis set, both $\overline{\Delta}$ and $\sigma$ are smaller than 1\%. In aug-cc-pVTZ, they are reduced by an order of magnitude. The results of the STG/C approach in aug-cc-pVDZ are not as stable as in STG/B with the same basis set---for three dimers (HF-methane, formaldehyde-ethene, and ethyne dimer) the relative errors exceed 1\%. Although in the STG/C approach $\overline{\Delta}$ is practically zero, in the aug-cc-pVTZ basis set the standard deviation is two times larger than in STG/B. It should be stressed, however, that the value $\sigma\approx 0.2\%$ obtained in STG/C is still of remarkable quality.

The dispersion dominated complexes pose a challenge for the explicitly correlated approach when the aug-cc-pVDZ basis set is employed. This is reflected in the large value of $\sigma$ for both STG/B and STG/C. Two of the class-D systems contain the argon atom---such dimers are known to be particularly difficult for F12 methods.~\cite{Patkowski:12,Patkowski:13}. Exclusion of these systems strongly affects the results: in aug-cc-pVTZ the values of both $\overline{\Delta}$ and $\sigma$ become similar to the ones observed for class-M complexes ($\overline{\Delta}=-0.05\%$ and $\sigma=0.21\%$ for STG/B; $\overline{\Delta}=-0.10\%$ and $\sigma=0.17\%$ for STG/C).


\subsection{Basis-set convergence for the water dimer}

\begin{figure}
\caption{\label{fig:water} Convergence of the dispersion energy for
the water dimer with the basis set size. The reference energy is $E^{(20)}_\mathrm{disp,CBS}=-10.933$ kJ/mol. Notice the $\log$-$\log$ scale.}
\includegraphics[width=\linewidth]{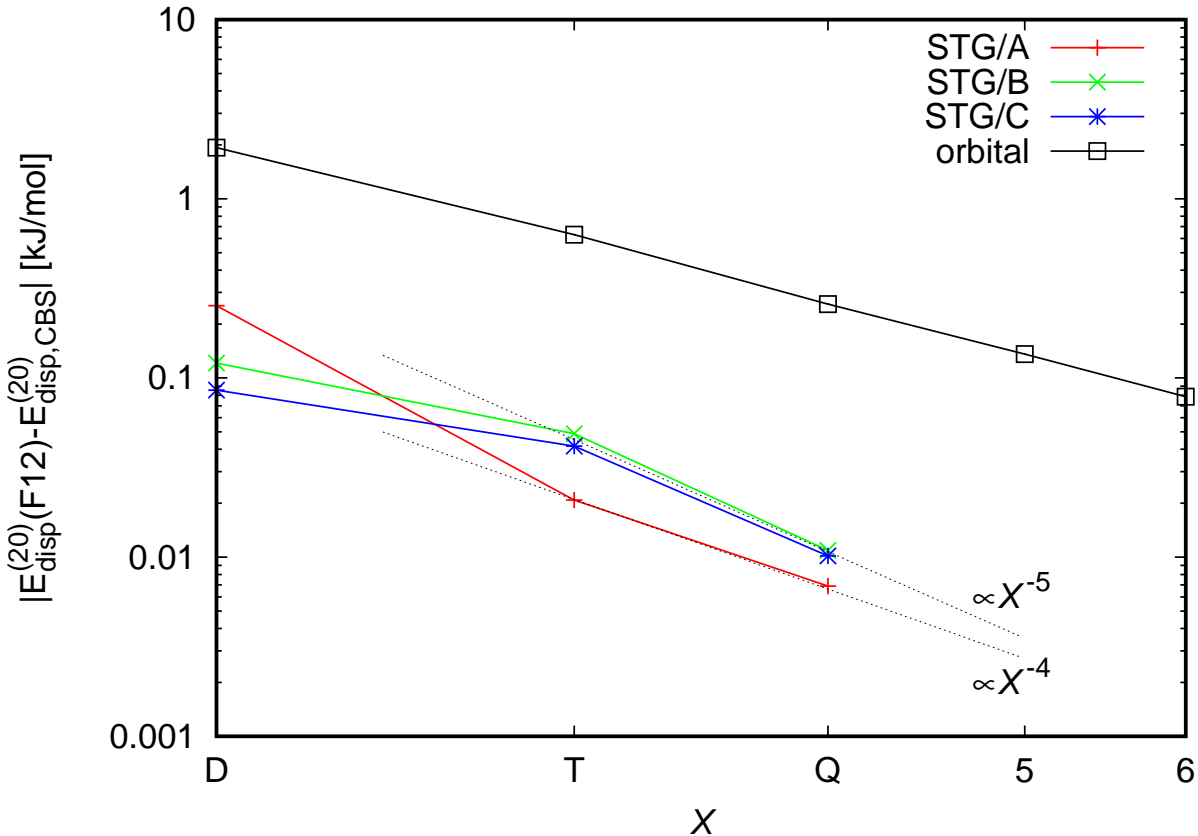}
\end{figure}

In Figure~\ref{fig:water} we present the basis set convergence of $E^{(20)}_\mathrm{disp}$(F12) for the water dimer. Calculations in both the explicitly correlated and standard approaches were performed in basis sets with the cardinal numbers $X$ up to $X\!=4$ and $X\!=6$, respectively, that is greater by one than in the previous sections. We show results for the STG correlation factor and for all variants of the F12 approach.

The STG/A dispersion energy converges with $X$ to the reference value from below, while both the STG/B and STG/C energies converge from above. This convergence pattern is consistent with previous observations for intramonomer correlation which led to the dispersion-weighted explicitly correlated approach.~\cite{Marshall:11} The error in the largest basis set is $-0.007$~kJ/mol for variant A and $0.011$~kJ/mol for both variants B and C.

Unlike in the orbital calculations, the convergence rate in the F12 approach is not constant. In variant A it decreases with the cardinal number $X$---the error is first reduced by an order of magnitude for $\mathrm{D}\rightarrow\mathrm{T}$, but for $\mathrm{T}\rightarrow\mathrm{Q}$ the reduction is only by a factor of 3. In contrast, in variants B and C the convergence rate increases---the error reduction by a factor of 2 for $\mathrm{D}\rightarrow\mathrm{T}$ is followed by a reduction by a factor of 4 for $\mathrm{T}\rightarrow\mathrm{Q}$. The power $n$ in the $X^{-n}$ convergence formula determined by a fit to the aug-cc-pVTZ and aug-cc-pVQZ results is $n=4$ in variant A and $n=5$ in both variants B and C. This convergence rate is faster than in the orbital calculations ($n=3$) but slower than the one derived from analytical studies for atoms ($n=5$ for variant A and $n=7$ for variant B).~\cite{Kutzelnigg:91} This is in line with the findings of Grant Hill \emph{et al.}~\cite{Hill:09} who analyzed basis set convergence of the MP2-F12 and CCSD-F12b correlation energies for molecules in the aug-cc-pV$X$Z sequence of basis sets.

\subsection{\label{sec:results:stretch}Large intermolecular distances}

\begin{figure*}
\caption{\label{fig:stretch} Dependence of the calculated $E^{(20)}_{\mathrm{disp}}$(F12) dispersion energy on the value of the length-scale parameter $\gamma$ for the water dimer ($2.0\,R_\mathrm{eq}$ geometry). All calculations employed the STG correlation factor. Solid and dashed lines represent values obtained with the aug-cc-pVDZ and aug-cc-pVTZ basis sets, respectively, with matching aug-cc-pV$X$Z\_OPTRI auxiliary basis sets. Horizontal line marks the $E^{(20)}_\mathrm{disp,CBS}$ reference.}
\includegraphics[width=\textwidth]{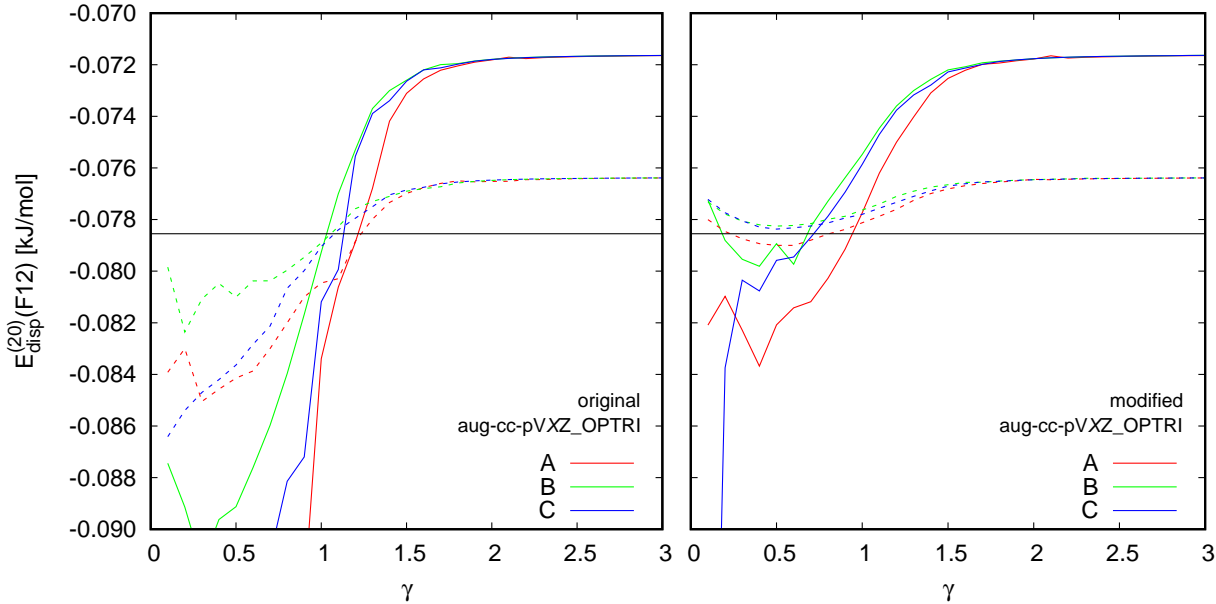}
\end{figure*}

To verify whether the good performance of the $E^{(20)}_\mathrm{disp}(\mathrm{F12})$ method for complexes in equilibrium geometries holds when the interacting molecules are far apart, we performed additional calculations in $1.5\,R_\mathrm{eq}$ and $2.0\,R_\mathrm{eq}$ geometries for three systems (water-water, water-methane, and methane-methane dimers). In the tests we used only the STG correlation factor. The representative results for the water dimer in the $2.0\,R_\mathrm{eq}$ geometry are presented in Figure~\ref{fig:stretch}. 

As can be inferred from the inspection of the left panel of Figure~\ref{fig:stretch}, the aug-cc-pVDZ\_OPTRI basis sets in their original form are inadequate at large intermolecular distances. For $\gamma<1.0$ the dispersion energies fall below the reference value irrespective of the F12 variant indicating that $E^{(20)}_{\mathrm{disp}}(\mathrm{F12})$ looses the variational character. This problem was already predicted by Yousaf and Peterson.~\cite{Yousaf:09} For weak noncovalent interactions they proposed extending the auxiliary basis sets by a few additional diffuse functions. In the present work we adapted a simple scheme where the exponents of the additional functions in aug-cc-pV$X$Z\_OPTRI were taken from a corresponding regular d-aug-cc-pV$X$Z basis set. For $l\leq l_\mathrm{max}$, where $l_\mathrm{max}$ is the maximal angular momentum number present in d-aug-cc-pV$X$Z, the smallest exponents from a given $l$-space of d-aug-cc-pV$X$Z were used. For $l> l_\mathrm{max}$, the additional exponents were fixed at the value of the smallest exponent from $l_\mathrm{max}$-space of d-aug-cc-pV$X$Z.

The results obtained with the modified version of the aug-cc-pV$X$Z\_OPTRI auxiliary basis sets are presented in the right panel of Figure~\ref{fig:stretch}. With the aug-cc-pVDZ basis set the largest improvement is seen for variant B. Still, in variants A and C the collapse of $E^{(20)}_{\mathrm{disp}}(\mathrm{F12})$ persists. Much better behavior is achieved in aug-cc-pVTZ. The shape of the $\gamma$-dependence of the dispersion energy is similar to that in Figure~\ref{fig:skan}. Both variants B and C exhibit a variational character, and the curves for all three variants have well-defined minima. Therefore, only the aug-cc-pVTZ basis set is used in further discussion.

The most important difference with respect to the equilibrium geometry is a shift of the optimal $\gamma$ to smaller values. For the water dimer $\gamma$ shifts to $\gamma=0.7$ and $\gamma=0.5$ for $1.5\,R_\mathrm{eq}$ and $2.0\,R_\mathrm{eq}$ geometries, respectively. For the water-methane and methane-methane dimers the shift is to $\gamma=0.5$ and $\gamma=0.4$ for $1.5\,R_\mathrm{eq}$ and $2.0\,R_\mathrm{eq}$. With optimal $\gamma$, the accuracy of the $E^{(20)}_\mathrm{disp}$(F12) method for larger intermolecular distances is similar to that observed for the equilibrium geometry. Version A overestimates the magnitude of the reference dispersion energies with average errors comparable to the results from the orbital calculations with aug-cc-pV5Z (0.5\% and 0.3\% for the $1.5\,R_\mathrm{eq}$ and $2.0\,R_\mathrm{eq}$ geometry, respectively). The average errors for both variants B and C are always smaller than 0.1\%.

\section{\label{sec:conclusions}Conclusions}

This work presents the first rigorous application of the modern F12 methodology for calculations of the second-order dispersion energy. We derived formulas for $E^{(20)}_\mathrm{disp}$(F12) in the A, B and C variants assuming the \textit{Ansatz}~\textbf{3} form of the strong orthogonality projector. We investigated three correlation factors: $r_{12}$, $\exp(\gamma\,r_{12})$, and $r_{12}\exp(-\gamma\,r_{12})$. The accuracy of the new approach was verified for the A24 data set of noncovalently bound dimers. As a reference the CBS-extrapolated dispersion energies from orbital-based calculations were used. 

For the exponential correlation factors we determined optimal values of the length-scale parameter $\gamma$ which guarantee highly accurate F12-dispersion energies irrespective of the character of the noncovalent interaction. In the van der Waals minimum region we recommend using $\gamma = 0.9$ for the STG correlation factor and $\gamma = 0.2$ for RSTG. 

Using optimal $\gamma$ values we found that the STG correlation factor yields most accurate dispersion energies, clearly superior to both R12- and RSTG-based results. When combined with either B or C variant the STG correlation factor leads to F12-dispersion energies of excellent quality. Already with aug-cc-pVDZ and aug-cc-pVTZ basis sets $E^{(20)}_\mathrm{disp}$(F12) reproduces the reference values with average percent error of the order of 0.1\%. The results obtained with aug-cc-pVTZ are more reliable as indicated by both the standard deviation and maximum absolute error being an order of magnitude smaller than in the double-zeta basis. The lower computational cost of the C approximation makes it a preferred choice for the F12-dispersion energy calculations.

For the A24 data set best results were achieved for mixed electrostatic/dispersion complexes with average relative errors smaller than 0.1\% with respect to the $E^{(20)}_{\rm disp,CBS}$ reference. For hydrogen bonded complexed we observed slight underestimation of the magnitude of the dispersion energy (average error of 0.3\%) resulting from a suboptimal character of the recommended value of $\gamma$ in this case. Results for the van der Waals complexes were overestimated by ca. 0.2\%, which we traced to large errors for Ar-methane and Ar-ethane dimers. When the argon atom containing systems are excluded from the analysis, the average relative errors are close to 0.1\%---similar to the level of accuracy achieved for mixed electrostatic/dispersion dimers. 

We stress that the average errors of $E^{(20)}_\mathrm{disp}$(F12) reported in this work are smaller than the degree of confidence of the orbital-based reference. The error of the reference dispersion energy, estimated as the difference between CBS-extrapolated results and results in the 5-zeta basis, is larger than 1\%. This indicates that explicitly correlated calculations of the dispersion energy have a potential to set a new standard for accurate predictions of this energy component.


Our preliminary results indicate that at large intermonomer separations the existing auxiliary basis sets are inadequate for the intermolecular correlation. In Section~\ref{sec:results:stretch} we showed that addition of diffuse functions is sufficient to circumvent this problem. In general, $E^{(20)}_\mathrm{disp}$(F12) calculations offer possibilities for designing auxiliary basis sets tailored to intermolecular interactions. Furthermore, our results revealed that the dependence of the optimal length-scale parameter on the intermolecular distance has to be accounted for to maintain high level of accuracy of the F12-dispersion energy. The character of this dependence as well as the possibility of improving the $E^{(20)}_\mathrm{disp}$(F12) behavior by employing correlation factor with the correct long-range asymptotics~\cite{Lesiuk:13} are being investigated in our group.

The first successful application of the F12 methodology to the second-order dispersion energy at the uncoupled level of theory reported in this work opens paths for further studies. First, it is worthwhile to investigate extension of the explicitly correlated approach to energy components involving intramonomer correlation, in particular $E^{(21)}_{\rm disp}$, and exchange components. Second, introduction of density-fitting techniques to F12-dispersion calculations, which we have not explored in this work, holds promise to broaden applications of the new method to medium and large systems which are already within the range of modern SAPT implementations.

\section{Acknowledgments}
The author would like to thank Dr. Micha\l{} Hapka for helpful discussions and commenting on the manuscript. 
This research was supported by National Science Centre (NCN) Grant No. 2012/05/D/ST4/01271.

\providecommand{\latin}[1]{#1}
\providecommand*\mcitethebibliography{\thebibliography}
\csname @ifundefined\endcsname{endmcitethebibliography}
  {\let\endmcitethebibliography\endthebibliography}{}

\end{document}